\documentclass{article}

\usepackage{arxiv}

\usepackage[utf8]{inputenc} 
\usepackage[T1]{fontenc}    
\usepackage{hyperref}       
\usepackage{url}            
\usepackage{booktabs}       
\usepackage{amsfonts}       
\usepackage{nicefrac}       
\usepackage{microtype}      
\usepackage{lipsum}		
\usepackage{graphicx}
\usepackage{natbib}
\usepackage{doi}
\usepackage{amsmath}

\graphicspath{{./figures/}}

\title{Nonlinear Dendritic Coincidence Detection for Supervised Learning}

\author{ \href{https://orcid.org/0000-0001-9156-0596}{\includegraphics[scale=0.06]{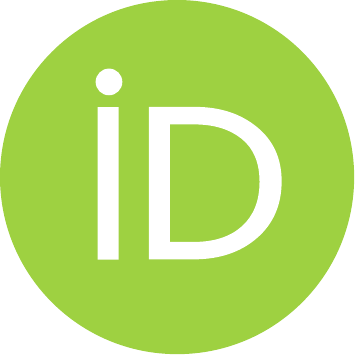}\hspace{1mm}Fabian~Schubert} \\
	Institute for Theoretical Physics\\
	Goethe University Frankfurt\\
	Frankfurt am Main\\
	\texttt{fschubert@itp.uni-frankfurt.de} \\
	\And
	\href{https://orcid.org/0000-0002-2126-0843}{\includegraphics[scale=0.06]{orcid.pdf}\hspace{1mm}Claudius~Gros} \\
	Institute for Theoretical Physics\\
	Goethe University Frankfurt\\
	Frankfurt am Main\\
	\texttt{gros@itp.uni-frankfurt.de} \\
}

\begin{document}
\maketitle

\begin{abstract}
	Cortical pyramidal neurons have a complex 
	dendritic anatomy, whose function is an active 
	research field. In particular, 
	the segregation between its soma and the apical 
	dendritic tree is believed to play an active 
	role in processing feed-forward sensory 
	information and top-down or feedback signals. 
	In this work, we use a simple two-compartment 
	model accounting for the nonlinear interactions 
	between basal and apical input streams and show 
	that standard unsupervised Hebbian learning rules 
	in the basal compartment allow the neuron to 
	align the feed-forward basal input with
	the top-down target signal received by the 
	apical compartment. 
	We show that this learning process, termed 
	coincidence detection, is robust against strong 
	distractions in the basal input space and 
	demonstrate its effectiveness in a linear 
	classification task.
\end{abstract}


\section{Introduction}
\label{sect:introduction}

In recent years, a growing body of research has addressed the 
functional implications of the distinct physiology and anatomy of 
cortical pyramidal neurons \citep{Spruston2008,Hay2011,Ramaswamy2015}. 
In particular, on the theoretical side,
we saw a paradigm shift from treating neurons as point-like electrical
structures towards embracing the entire dendritic structure 
\citep{Larkum2009,Poirazi2009,Shai2015}. This was 
mostly due to the fact that experimental work 
uncovered dynamical properties of pyramidal neuronal
cells that simply could not be accounted for by point models
\citep{Spruston1995,Hausser2000}.

An important finding is that the apical dendritic tree of
cortical pyramidal neurons can act as a separate nonlinear synaptic 
integration zone \citep{Spruston2008,Branco2011}. 
Under certain conditions, a dendritic $\rm Ca^{2+}$ spike
can be elicited that propagates towards the soma, causing rapid, bursting
spiking activity. One of the cases in which dendritic spiking can occur
was termed `backpropagation-activated $\rm Ca^{2+}$ spike firing' 
(`BAC firing'): A single somatic spike can backpropagate towards the apical
spike initiation zone, in turn significantly facilitating the initiation of 
a dendritic spike \citep{Stuart2001,Spruston2008,Larkum2013}. 
This reciprocal coupling is believed to act as a form of
coincidence detection: If apical and basal synaptic input co-occurs, the 
neuron can respond with a rapid burst of spiking activity. 
The firing rate of these temporal bursts exceeds the firing 
rate that is maximally achievable under basal synaptic input alone, 
therefore representing a form of temporal coincidence
detection between apical and basal input.

Naturally, these mechanisms also affect plasticity, and thus learning
within the cortex \citep{Sjoestroem2006,Ebner2019}. 
While the interplay between basal and apical stimulation and
its effect on synaptic efficacies is subject to ongoing research, 
there is evidence that BAC-firing tends to shift plasticity 
towards long-term potentiation (LTP) \citep{Letzkus2006}. 
Thus, coincidence between basal and apical input appears 
to also gate synaptic plasticity.

In a supervised learning scheme, where the top-down input
arriving at the apical compartment acts as the teaching signal,
the most straight-forward learning rule for the basal synaptic
weights would be derived from an appropriate loss function,
such as a mean square error, based on the difference between 
basal and apical input, i.e.\ $I_p - I_d$,
where indices $p$ and $d$ denote `proximal' and
`distal', in equivalence to basal and apical. 
Theoretical studies have investigated possible 
learning mechanisms that could utilize an 
intracellular error signal
\citep{Urbanczik2014,Schiess2016,Guerguiev2017}.
However, a clear experimental
evidence for a physical quantity encoding such an error 
is---to our knowledge---yet to be found. 
On the other hand, Hebbian-type plasticity is extensively
documented in experiments 
\citep{Gustafsson1987,Debanne1994,Markram1997,Bi1998}. 
Therefore, our work is based on the question of whether the 
nonlinear interactions between basal and apical synaptic input could, 
when combined with a Hebbian plasticity rule, allow a neuron
to learn to reproduce an apical teaching signal in its
proximal input.

We investigate coincidence learning by
combining a phenomenological model that 
generates the output firing rate as a function 
of two streams of synaptic input (subsuming basal 
and apical inputs) with classical Hebbian, as well as 
BCM-like plasticity rules on basal synapses. 
In particular, we hypothesized that this combination of neural 
activation and plasticity rules would lead to an
increased correlation between basal and apical inputs.
Furthermore, the temporal alignment observed in our study 
could potentially facilitate apical inputs to act as 
top-down teaching signals, without the need for an 
explicit error-driven learning rule. Thus, we also 
test our model in a simple linear supervised 
classification task and compare it with the 
performance of a simple point neuron equipped with 
similar plasticity rules.

\section{Model}
\label{sect:model}

\subsection{Compartamental Neuron}
\label{sect:neuronmodel}

The neuron model used throughout this study 
is a discrete-time rate encoding model that 
contains two separate input variables, 
representing the total synaptic input current injected arriving 
at the basal (proximal) and apical (distal) 
dendritic structure of a pyramidal neuron, respectively. 
The model is a slightly simplified version of a phenomenological 
model proposed by \citet{Shai2015}. Denoting the input currents 
$I_p$ (proximal) and $I_d$ (distal), 
the model is written as
\begin{align}
\begin{split}
y\left(t\right) &= \alpha  \sigma\left( I_p(t) - \theta_{p0} \right)
\left[1-\sigma\left(I_d(t) - \theta_d\right)\right] \\
&+ \sigma\left(I_d(t) - \theta_d \right)
\sigma\left( I_p(t) - \theta_{p1} \right)
\end{split} 
\label{eq_comp_model}\\
\sigma(x) &\equiv \frac{1}{1+\exp(-4x)} \; .
\end{align}
Here, $\theta_{p0}>\theta_{p1}$ and $\theta_d$ 
are threshold variables with respect to proximal 
and distal inputs. 
Equation (\ref{eq_comp_model}) defines the firing
rate $y$ as a function of $I_p$ and $I_d$. Note that
the firing rate is normalized to take values within
$y \in [0,1]$. In the publication by \citet{Shai2015},
firing rates varied between $0$ and $150\,\mathrm{Hz}$.
High firing rates typically appear in the form of bursts
of action potentials, lasting on the order of $50$--$100 \, \mathrm{ms}$
\cite{Larkum1999,Shai2015}. Therefore, since our model represents
``instantaneous" firing rate responses to a discrete set of static
input patterns, we conservatively estimate the time scale
of our model to be on the order of tenths of seconds.

In general, the input currents $I_p$ and $I_d$ are meant
to comprise both excitatory and potential inhibitory
currents. Therefore, we did not restrict the sign of
of $I_p$ and $I_d$ to positive values. Moreover, since
we chose the thresholds $\theta_{p0}$ and $\theta_{d}$
to be zero, $I_p$ and $I_d$ should be rather seen as
a total external input relative to intrinsic firing
thresholds.

Note that the original form of this phenomenological
model by \citet{Shai2015} is of the form
\begin{equation}
y(I_p, I_d) = \sigma\left( I_p - A \sigma(I_d) \right)\left[1 + B \sigma(I_d)\right] \; ,
\end{equation}
where $\sigma$ denotes the same sigmoidal activation function. This equation
illustrates that $I_d$ has two effects: It shifts the basal activation threshold
by a certain amount (here controlled by the parameter $A$) and also multiplicatively
increases the maximal firing rate (to an extent controlled by $B$). Our equation
mimics these effects by means of the two thresholds $\theta_{p0}$ and $\theta_{p1}$, as well
as the value of $\alpha$ relative to the maximal value of y (which is $1$ in our case).

\begin{figure}[t]
	\centering
	\includegraphics[width=0.6\columnwidth]{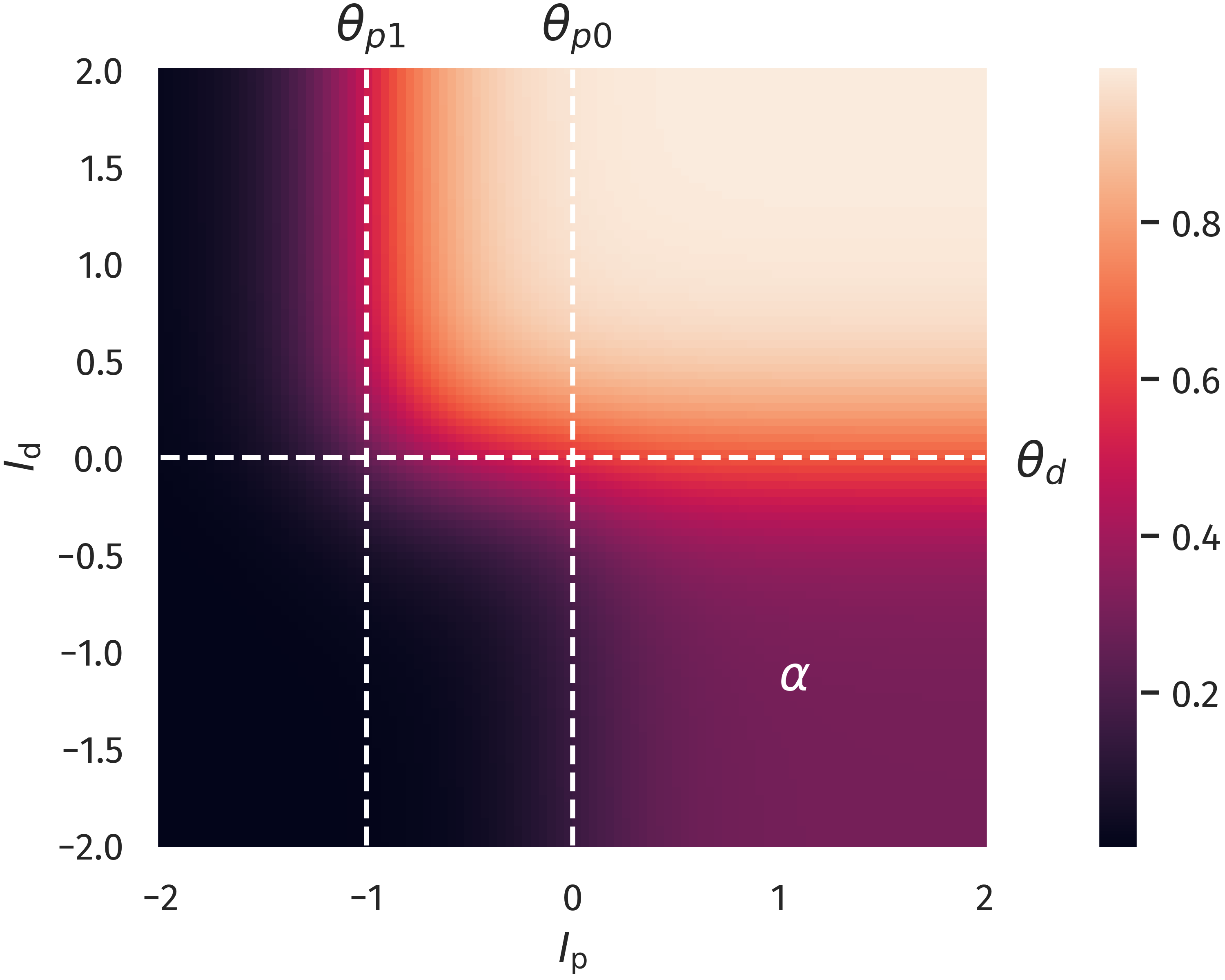}
	\caption{{\bf Two-compartment rate model.} 
		The firing rate as a function of proximal and distal 
		inputs $I_p$ and $I_d$, 
		see \eqref{eq_comp_model}. 
		The thresholds $\theta_{p0}$, $\theta_{p1}$ and 
		$\theta_d$ define two regions of neural activity,
		with a maximal firing rate of $1$ and a plateau
		in the lower-left quadrant with a value of $\alpha=0.3$. That is,
		the latter region can achieve $30\%$ of the maximal firing rate.
	}
	\label{fig_comp_model}
\end{figure}

Overall, equation 
(\ref{eq_comp_model}) describes two distinct
regions of neural activation in the 
$(I_p, I_d)$-space which differ in their
maximal firing rates, which are set to $1$ and 
$\alpha$, where $0 < \alpha < 1$.
A plot of \eqref{eq_comp_model} is shown 
in Fig.~\ref{fig_comp_model}.

When both input currents $I_d$ and $I_p$ 
are large, that is, larger than the 
thresholds $\theta_d$ and $\theta_{p1}$,
the second term in \eqref{eq_comp_model}
dominates, which leads to $y\approx 1$. 
An intermediate activity plateau, of
strength $\alpha$ emerges in addition 
when $I_p>\theta_{p0}$ and 
$I_d<\theta_{d}$. As such, the compartment
model \eqref{eq_comp_model} is able to
distinguish neurons with a normal activity 
level, here encoded by $\alpha=0.3$, and
strongly bursting neurons, where the maximal
firing rate is unity. The intermediate plateau
allows neurons to process the proximal 
inputs $I_p$ even in the absence of distal
stimulation. The distal current $I_d$
acts therefore as an additional modulator.

In our numerical experiments, we compare 
the compartment model with a classical point 
neuron, as given by
\begin{equation}
y(t) = \sigma\left(I_p(t) + I_d(t) - 
\theta \right) \; .
\label{eq_point_neuron}
\end{equation}
The apical input $I_d$ is generated
`as is', meaning it is not dynamically 
calculated as a superposition of multiple 
presynaptic inputs. For concreteness, we
used
\begin{equation}
I_d(t) = n_d(t) x_d(t) - b_d(t) \; ,
\label{eq_I_d}
\end{equation}
where $n_d(t)$ is a scaling factor, $x_d(t)$ 
a discrete time sequence, which represents the target
signal to be predicted by the proximal input, and 
$b_d(t)$ a bias. In our experiments, we chose $x_d$ 
according to the prediction task at hand, see 
(\ref{eq_x_d_a}) and (\ref{eq:bin_targ_0})--(\ref{eq:bin_targ_1}).

Note that $n_d$ and $b_d$ 
are time dependent since they are subject 
to adaptation processes, which will be
described in the next section. Similarly, 
the proximal input $I_p(t)$ is given by
\begin{equation}
I_p(t) = n_p(t) \sum_{i=1}^{N} 
x_{p,i}(t) w_i(t) - b_p(t) \; ,
\label{eq_I_p}
\end{equation}
where $N$ is the number of presynaptic afferents, 
$x_{p,i}(t)$ the corresponding sequences, 
$w_i(t)$ the synaptic efficacies and
$n_p(t)$ and $b_p(t)$ the (time dependent)
scaling and bias. Tyical values for the 
parameters used throughout this study
are presented in Table~\ref{tab_parameters}.

\subsection{Homeostatic Parameter Regulation\label{sect_homeostasis}}

The bias variables entering the definitions
\eqref{eq_I_d} and \eqref{eq_I_p} of the distal
proximal current, $I_d$ and $I_p$, 
are assumed to adapt according to
\begin{align}
\label{eq_b_p_dot}
b_p(t+1) &= b_p(t) + \mu_b \left[I_p(t) - I_p^t\right] \\
b_d(t+1) &= b_d(t) + \mu_b \left[I_d(t) - I_d^t\right] \;,
\label{eq_b_d_dot}
\end{align}
where $I_p^t=0$ and, $I_d^t=0$
are preset targets and $1/\mu_b=10^3$ is the
timescale for the adaption. 
Since this is a slow process, over time, 
both the distal and the proximal 
currents, $I_d$ and $I_p$, will approach a temporal mean
equal to $I_p^t$ and $I_d^t$ respectively while still allowing the input
to fluctuate.
The reason for choosing  the targets to be
zero lies in the fact that we expect a neuron
to operate in a dynamical regime that can reliably
encode information from its inputs. In the
case of our model, this implies that neural
input should be distributed close to the threshold (which
was set to zero in our case),
such that fluctuations in the can have an effect
on the resulting neural activity. See e.g.\ \citet{Bell_1995} and
\citet{Triesch_2007} for theoretical approaches to 
optimizing gains and biases based on input and output statistics.
Hence, while we chose the mean targets of the input to be
the same as the thresholds, this is not a strict condition,
as relevant information in the input could also be present in
parts of the input statistics that significantly differ from its
actual mean (for example in the case of a heavily skewed distribution).

Adaptation rules for the bias entering a 
transfer function, such as
\eqref{eq_b_d_dot} and \eqref{eq_b_p_dot},
have the task to regulate overall activity
levels. The overall magnitude of the
synaptic weights, which are determined by 
synaptic rescaling factors, here $n_d$ and $n_p$, 
as defined in \eqref{eq_I_d} and \eqref{eq_I_p},
will regulate in contrast the variance of
the neural activity, and not the average
level \citep{schubert2021local}. In this
spirit we consider
\begin{align}
n_d(t+1) &= n_d(t) + 
\mu_n \left[V_d^t - \left( I_d(t) - \tilde{I}_d(t)\right)^2\right] \\
n_p(t+1) &= n_p(t) + 
\mu_n \left[V_p^t - \left( I_p(t) - \tilde{I}_p(t)\right)^2\right] \\
\tilde{I}_d(t+1) &= (1-\mu_{\rm av})\tilde{I}_d(t) + \mu_{\rm av} I_d(t) \label{eq:gain_d} \\
\tilde{I}_p(t+1) &= (1-\mu_{\rm av})\tilde{I}_p(t) + \mu_{\rm av} I_p(t) \; . \label{eq:gain_p}
\end{align}
Here, $V_p^t$ and $V_p^t$ define targets for 
the temporally averaged variances of $I_p$ 
and $I_d$.  
The dynamic variables $\tilde{I}_p$ and $\tilde{I}_d$ 
are simply low-pass filtered running averages of 
$I_p$ and $I_d$. Overall, the framework
specified here allows the neuron to be fully flexible,
as long as the activity level and its variance 
fluctuate around preset target values 
\citep{schubert2021local}. 

Mapping the control of the mean input current to the
biases and the control of variance to the gains is, in a sense,
an idealized case of the more general notion of dual homeostasis.
As shown by \citet{cannon2017stable}, the conditions for 
a successful control of mean and variance by means of gains and
biases are relatively loose: Under certain stability conditions,
a combination of two nonlinear functions of the variable that is
to be controlled can yield a dynamic fixed point associated with
a certain mean and variance. In fact, a possible variant of dual
homeostasis could potentially be achieved by coupling the input
gains to a certain firing rate (which is a non-linear function
of the input), while biases are still adjusted to a certain mean
input. This, of course, would make it harder to predict the
variance of the input resulting from such an adaptation, since
it would not enter the equations as a simple parameter that can
be chosen a priori (as it is the case for equation (\ref{eq:gain_d}) and
(\ref{eq:gain_p})).

A list of 
the parameter values used throughout this investigation 
is also given in Table~\ref{tab_parameters}. Our choices of
target means and variances are based on the assumption
that neural input should be tuned towards a certain
working regime of the neural transfer function. In
the case of the presented model, this means that both
proximal and distal input cover an area where the nonlinearities
of the transfer function are reflected without oversaturation.

\begin{table}[b]
	\caption{Model parameters,
		as defined in sections \ref{sect:neuronmodel} and
		\ref{sect_plasticity}.}
	\begin{tabular}{ l | l || l | l }
		$\theta_{p0}$ & $0$ & $V_d^t$ & $0.25$ \\
		$\theta_{p1}$ & $-1$ & $\mu_b$ & $10^{-3}$ \\ 
		$\theta_{d}$ & $0$ & $\mu_n$ & $10^{-4}$ \\  
		$\alpha$ & $0.3$ & $\mu_{\rm av}$ & $5 \cdot 10^{-3}$ \\   
		$\mu_w$ & $5 \cdot 10^{-5}\quad$ & $I_p^t$ & $0$ \\
		$\epsilon$ & $0.1$ & $I_d^t$ & $0$ \\
		$V_p^t$ & $0.25$ &  &
	\end{tabular}
	\label{tab_parameters}
\end{table}

\subsection{Synaptic Plasticity\label{sect_plasticity}}

The standard Hebbian plasticity rule for the 
proximal synaptic weights is given by
\begin{align}
w_i(t+1) &= w_i(t) + \mu_w 
\big[\left(x_{p,i}(t) - \tilde{x}_{p,i}(t)\right)
\left(y(t) - \tilde{y} \right) - \epsilon w_i(t) \big]
\label{eq_hebb_plast} \\
\tilde{x}_{p,i}(t+1) &= 
(1-\mu_{\rm av})\tilde{x}_{p,i}(t) + \mu_{\rm av}x_{p,i}(t) 
\label{eq_tilde_x_p_i}\\
\tilde{y}(t+1) &= (1-\mu_{\rm av})\tilde{y}(t) + \mu_{\rm av}y(t)
\label{eq_tilde_y}
\end{align}
The trailing time averages $\tilde{x}_{p,i}$ and
$\tilde{y}$, respectively of the presynaptic 
basal activities, $x_{p,i}$, and of the neural
firing rate $y$, enter the Hebbian learning
rule (\ref{eq_hebb_plast}) as reference levels.
Pre- and post-synaptic neurons are considered
to be active/inactive when being above/below
the respective trailing averages. 
This is a realization of the Hebbian rule
proposed by \citet{Linsker1986}.
The timescale of the averaging, $1/\mu_{\rm av}$,
is 200 time steps, see
Table~\ref{tab_parameters}. As discussed in 
Section \ref{sect:neuronmodel}, a time step can be
considered to be on the order of $100 \, \mathrm{ms}$, which
equates to an averaging time of about $20 \, \mathrm{s}$.
Generally, this is much faster than the timescales
on which metaplasticity, i.e.\, adaptation processes
affecting the dynamics of synaptic plasticity itself,
are believed to take place, which are on the
order of days \citep{Yger2015}. However, it should
be noted that our choice of the timescale of the 
averaging process used in our plasticity model is 
motivated mostly by considerations regarding the overall 
simulation time: Given enough update steps, the same
results could be achieved by an arbitrarily slow
averaging process.

Since classical Hebbian learning does not keep
weights bounded, we use an additional proportional decay
term $\epsilon w_i$ which prevents runaway
growth using $\epsilon = 0.1$.
With $1/\mu_w=2\cdot10^4$, learning is 
assumed to be considerably slower, 
as usual for statistical update rules.
For comparative reasons, the point neuron 
model (\ref{eq_point_neuron}) is equipped 
with the same plasticity rule for the proximal 
weights as (\ref{eq_hebb_plast}).

Apart from classical Hebbian learning, we also considered 
a BCM-like learning rule for the basal weights 
\citep{Bienenstock1982,Intrator1992}.
The form of the BCM-rule used here reads
\begin{equation}
w_i(t+1) = w_i(t) + \mu_w \big[ y\left(y - \theta_M\right) x_i - 
\epsilon w_i \big] \; , 
\label{eq_bcm_rule}
\end{equation}
where $\theta_M$ is a threshold defining a 
transition from long-term potentiation (LTP) to long-term depression (LTD) and, 
again, $\epsilon$ 
is a decay term on the weights preventing unbounded 
growth. In the variant introduced by \citet{Law1994}, 
the sliding threshold is simply the temporal average 
of the squared neural activity, 
$\theta_M = \langle y^2 \rangle$. In practice, 
this would be calculated as a running average, 
thereby preventing the weights from growing 
indefinitely.

However, for our compartment model, we chose to explicitly set the
threshold to be the mean value between the high- and low-activity regime
in our compartment model, i.e.\ $\theta_M = (1+\alpha)/2$. By doing so, LTP is
preferably induced if both basal and apical input is present at the same
time.
Obviously, for the point model, the reasoning behind our choice of
$\theta_M$ did not apply. Still, to provide some level of comparability,
we also ran simulations with a point model where the sliding threshold was
calculated as a running average of $y^2$.

\begin{figure}[t]
	\centering
	\includegraphics[width=0.55\columnwidth]{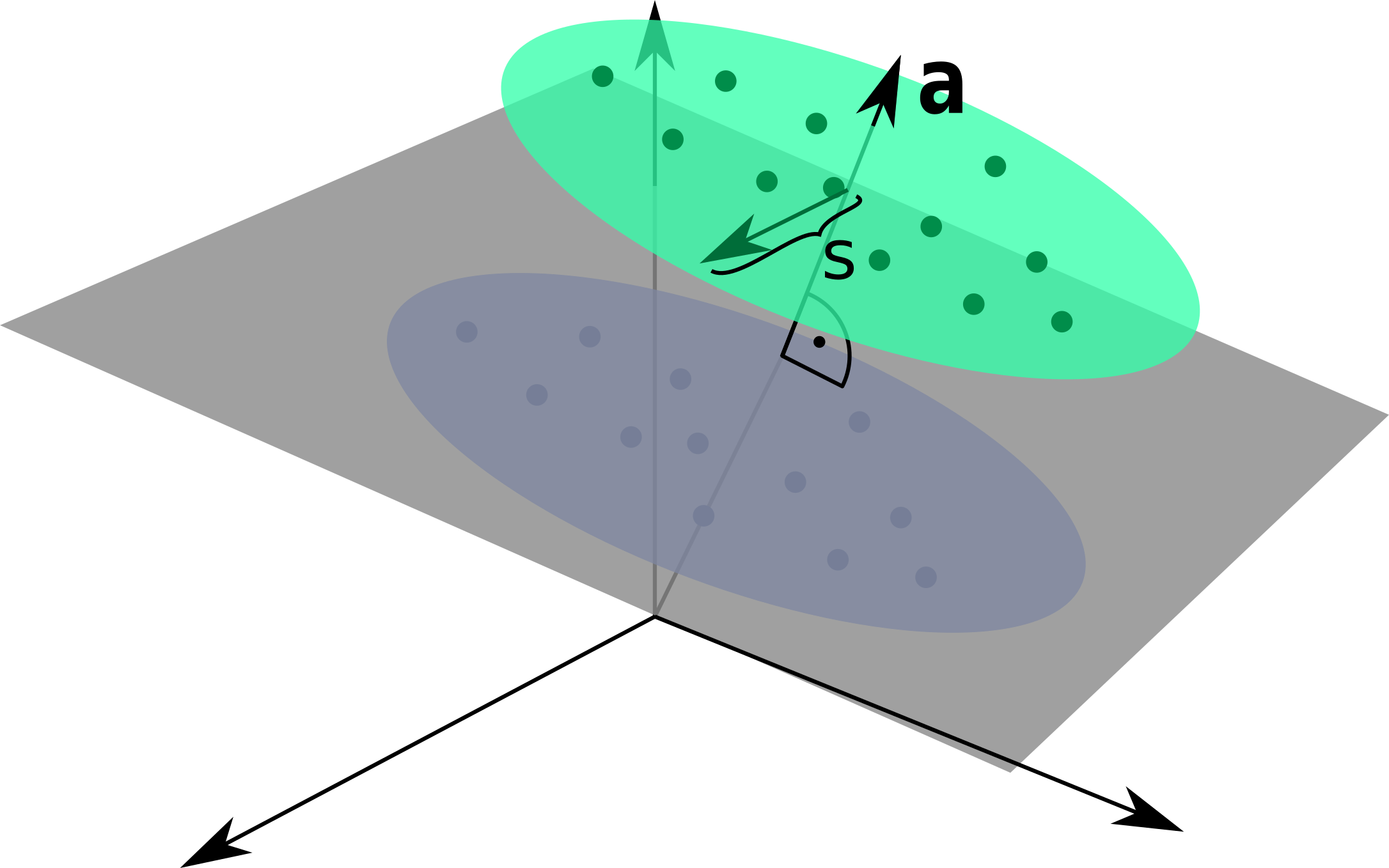}
	\caption{{\bf Input Space for the Linear Classification Task.}
		Two clusters of presynaptic basal activities were generated from 
		multivariate Gaussian distributions. Here, $s$ denotes the standard
		deviation orthogonal to the normal vector $\mathbf{a}$ of the 
		classification hyperplane, as defined by \eqref{eq_x_d_a}.}
	\label{fig_illustration_classification}
\end{figure}

\section{Results}
\label{sect:results}

\subsection{Unsupervised Alignment between Basal and Apical Inputs}
\label{sect:alignment}

As a first test, we quantify the neuron's ability to 
align its basal input to the apical teaching signal.
This can be done using the Pearson correlation coefficient
$\rho[I_p,I_d]$ between the basal and 
apical input currents. We determined 
$\rho[I_p,I_d]$ after the simulation,
which involves all plasticity mechanisms, both
for the synaptic weights and the intrinsic 
parameters. The input sequences $x_{p,i}(t)$ 
is randomly drawn from a uniform distribution,
in $[0,1]$, which is done independently for 
each $i\in[1,N]$.

For the distal current $I_d(t)$ to be fully 
`reconstructable' by the basal input, $x_d(t)$ has 
to be a linear combination 
\begin{align}
x_d(t) &= \sum_{i=1}^N a_i x_{p,i}(t)
\label{eq_x_d_a}
\end{align}
of the $x_{p,i}(t)$, where the $a_i$ are the
components of a random vector $\mathbf{a}$ 
of unit length. 

Given that we use with \eqref{eq_hebb_plast}
a Hebbian learning scheme, one can expect that
the direction and the magnitude of the principal 
components of the basal input may affect the
outcome of the simulation significantly:
A large variance in the basal input 
orthogonal to the `reconstruction vector' $\mathbf{a}$ 
is a distraction for the plasticity. The
observed temporal alignment between $I_p$ and 
$I_d$ should hence suffer when such a 
distraction is present. 

In order to test the effects of distracting directions,
we applied a transformation to the input sequences $x_{p,i}(t)$.
For the transformation, two parameters are used,
a scaling factor $s$ and the dimension $N_{\rm dist}$ 
of the distracting subspace within the basal input 
space. The $N_{\rm dist}$ randomly generated
basis vectors are orthogonal to the superposition
vector $\mathbf{a}$, as defined by \eqref{eq_x_d_a},
and to each others. Within this $N_{\rm dist}$-dimensional 
subspace, the input sequences $x_{p,i}(t)$ are
rescaled subsequently by the factor $s$. 
After the learning phase, a second set of input 
sequences $x_{p,i}(t)$ and $x_d(t)$ is generated for 
testing purposes, using the identical protocol, and
the cross correlation $\rho[I_p,I_d]$ 
evaluated. During the testing phase plasticity is 
turned off.

\begin{figure}[t]
	\centering
	\includegraphics[width=1.0\columnwidth]{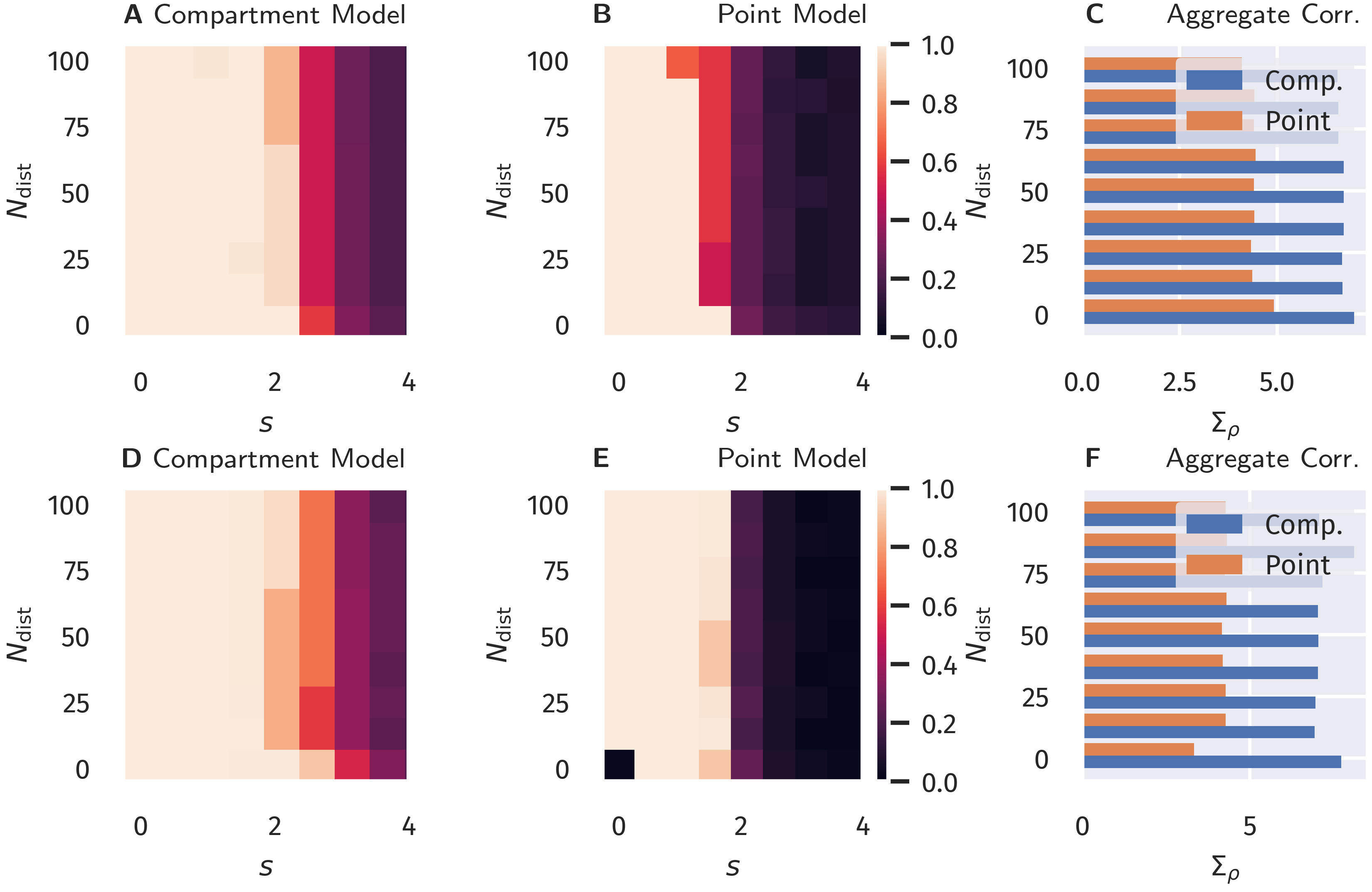}
	\caption{{\bf Unsupervised Alignment between 
			Basal and Apical Input.} Color
		encoded is the Pearson correlation $\rho[I_p,I_d]$ 
		between the proximal and distal input currents,
		$I_p$ and $I_d$. A--C: Classical Hebbian 
		plasticity, as defined by \eqref{eq_hebb_plast}.
		D--F: BCM rule, see \eqref{eq_bcm_rule}. 
		Data for a range $N_{\rm dist}\in[0,N-1]$
		of the orthogonal distraction directions, 
		and scaling factors $s$, as defined in
		Fig.~\ref{fig_illustration_classification}.
		The overall number of basal inputs is $N=100$.
		In the bar plot on the right the sum
		$\Sigma_{\rm acc}$ over $s=0,\,0.5,\,1.0\,..$ 
		of the results is shown as a function
		of $N_{\rm dist}$. Blue bars represents 
		the compartment model, orange the point model.}
	\label{fig_corr_dimension_scaling}
\end{figure}

The overall aim of our protocol is to evaluate
the degree $\rho[I_p,I_d]$ to which
the proximal current $I_p$ aligns in the
temporal domain to the distal input $I_d$.
We recall that this is a highly non-trivial
question, given that the proximal synaptic
weights are adapted via Hebbian plasticity, 
see \eqref{eq_hebb_plast}. The error 
$(I_p-I_d)^2$ does not
enter the adaption rules employed.
Results are presented in
Fig.~\ref{fig_corr_dimension_scaling}
as a function of the distraction parameters 
$s$ and $N_{\rm dist}\in[0,N-1]$. The
total number of basal inputs is $N=100$.

For comparison, in 
Fig.~\ref{fig_corr_dimension_scaling}
data for both the compartment model
and for a point neuron are presented
(as defined respectively by \eqref{eq_comp_model}
and \eqref{eq_point_neuron}), as well as results for both
classical Hebbian and BCM learning rules. A decorrelation transition as a function of the distraction scaling parameter $s$ is observed for both models and plasticity rules. In terms of the learning rules,
only marginal differences are present. 
However, the compartment model is able
to handle a significantly stronger distraction 
as compared to the point model. These findings
support the hypothesis examined here, namely
that nonlinear interactions between basal and 
apical input improve learning guided by top-down signals.

\subsection{Supervised Learning in a Linear Classification Task}
\label{sect:classification}

Next, we investigated if the observed differences would also improve
the performance in an actual supervised learning task.
For this purpose, we constructed presynaptic basal 
input $x_p(t)$ as illustrated in 
Fig.~\ref{fig_illustration_classification}.
Written in vector form, each sample from the basal 
input is generated from,
\begin{equation}
\mathbf{x}_p(t) = \mathbf{b} + 
\mathbf{a}\big[c(t) + \sigma_a \zeta_a(t) \big] 
+ s \cdot \sum_{i=1}^{N_{\rm dist}} 
\zeta_{dist,i}(t) \mathbf{v}_{{\rm dist},i} \; ,
\end{equation}
where $\mathbf{b}$ is a random vector, where each entry is drawn uniformly from
$[0,1]$, $\mathbf{a}$ is random unit vector as introduced in 
Section~\ref{sect:alignment}, $c(t)$ is a binary variable drawn 
from $\{-0.5,0.5\}$ with equal probability and $\zeta_a(t)$ and the
$\zeta_{dist,i}(t)$ are independent Gaussian random variables with 
zero mean and unit variance. 
Hence, $\sigma_a$ simply denotes the standard deviation of each Gaussian
cluster along the direction of the normal vector $\mathbf{a}$ and was
set to $\sigma_a = 0.25$. 
Finally, the set of $\mathbf{v}_{{\rm dist},i}$ forms a 
randomly generated orthogonal basis of $N_{\rm dist}$ 
unit vectors which are---as in 
Section~\ref{sect:alignment}---also orthogonal to 
$\mathbf{a}$. The free parameter $s$ parameterizes
the standard deviation along this subspace orthogonal 
to $\mathbf{a}$. As indicated by the time dependence, 
the Gaussian and binary random variables are drawn 
for each time step. The vectors $\mathbf{b}$, 
$\mathbf{a}$, and $\mathbf{v}_{{\rm dist},i}$ are generated
once before the beginning of a simulation run.

For the classification task, we use two output 
neurons, indexed 0 and 1,  receiving the same 
basal presynaptic input, with the respective 
top-down inputs $x_{d,0}$ and $x_{d,1}$ encoding 
the desired linear classification in a one-hot 
scheme, 
\begin{align}
x_{d,0}(t) &= 1 - \Theta\left( \left(\mathbf{x}_p(t) - 
\mathbf{b}\right)^T \mathbf{a}\right) \label{eq:bin_targ_0}\\
x_{d,1}(t) &= \Theta\left( \left(\mathbf{x}_p(t) - 
\mathbf{b}\right)^T \mathbf{a}\right) \; , \label{eq:bin_targ_1}
\end{align}
where $\Theta(x)$ is the Heaviside step function.

As in the previous experiment, we ran a full simulation until all dynamic variables reached a stationary state. After this,
a test run without plasticity and with the apical input turned off 
was used to evaluate the classification performance. 
For each sample, the index of the neuron with the highest 
activity was used as the predicted class. 
Accuracy was then calculated as the fraction
of correctly classified samples.
\begin{figure}[t]
	\centering
	\includegraphics[width=1.0\columnwidth]{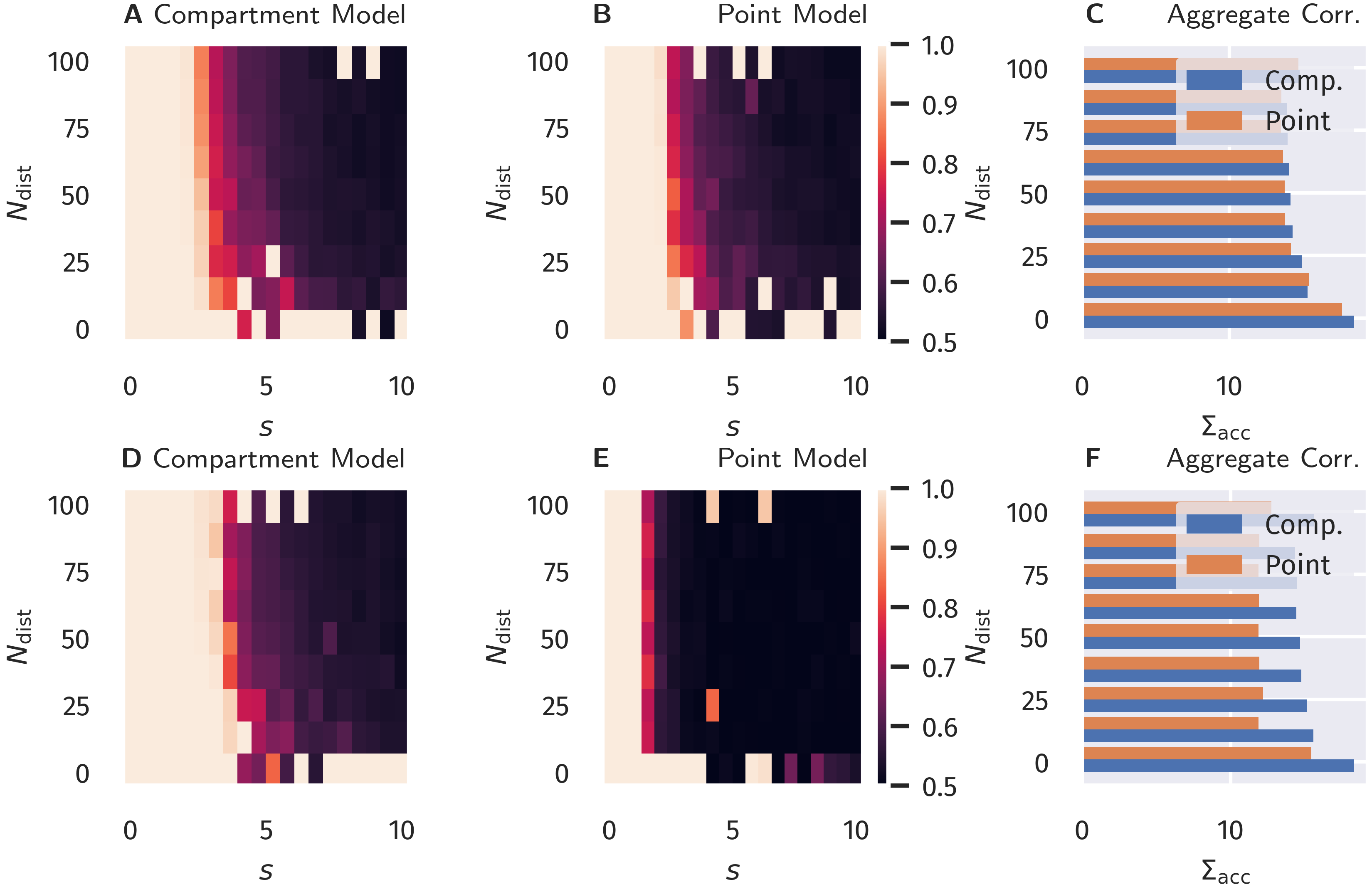}
	\caption{{\bf Binary Classification Accuracy.}
		Fraction of correctly classified patterns as illustrated in
		Fig.~\ref{fig_illustration_classification}, see 
		Section~\ref{sect:classification}. 
		A--C: Classical Hebbian plasticity. 
		D--F: BCM rule.
		In the bar plot on the right the sum 
		$\Sigma_{\rm acc}$ over $s=0,\,0.5,\,1.0\,..$ 
		of the results is given as a function
		of $N_{\rm dist}$. Blue bars 
		represents the compartment model,
		orange the point model.}
	\label{fig:classification_accuracy}
\end{figure}

The resulting accuracy as a function of $N_{\rm dist}$ and $s$
is shown in Fig.~\ref{fig:classification_accuracy}, again for all
four combinations of neuron models and learning rules.

For classical Hebbian plasticity, the differences between
compartmental and point neuron are small. Interestingly,
the compartment model performs measurably better in
the case of the BCM rule (\ref{eq_bcm_rule}), in 
particular when the overall accuracies for the 
tested parameter range are compared, see 
Fig.~\ref{fig:classification_accuracy}D.
This indicates that during learning, the compartmental neuron
makes better use,  of the three distinct activity plateaus at
$0$, $\alpha$ and $1$, when the BCM rule is at work. Compare
Fig.~\ref{fig_comp_model}. We point
out in this respect that the sliding 
threshold $\theta_M$ in (\ref{eq_bcm_rule}) 
has been set to the point halfway between
the two non-trivial activity levels,
$\alpha$ and $1$. 

It should be noted that the advantage of the 
compartment model is also reflected in the actual
correlation between proximal and distal input 
as a measure of successful learning (as done 
in the previous section), see 
Fig.~\ref{fig:classification_correlation} in 
the appendix. Interestingly,
the discrepancies are more pronounced when measuring
the correlation as compared to the accuracy. Moreover,
it appears that above-chance accuracy is still present
for parameter values where alignment is almost zero.
We attribute this effect to the fact that the 
classification procedure predicts the class by
choosing the node that has the higher activity, independent
of the actual ``confidence" of this prediction, i.e.\, how strong
activities differ relative to their actual activity levels.
Therefore, marginal differences can still
yield the correct classification in this isolated setup, 
but it would be easily disrupted by finite levels of noise or
additional external input.

\section{Discussion}

Pyramidal neurons in the brain
possess distinct apical/basal (distant/proximal)
dendritic trees. It is hence likely that models
with at least two compartments are necessary for
describing the functionality of pyramidal neurons.
For a proposed two-compartment transfer function
\citep{Shai2015},
we have introduced both unsupervised and supervised 
learning schemes, showing that the two-compartment
neuron is significantly more robust against 
distracting components in the proximal input 
space than a corresponding (one-compartment) point
neuron.

The apical and basal dendritic compartments of
pyramidal neurons are located in different
cortical layers \cite{park2019contribution},
receiving top-down and feed-forward
signals, respectively. The combined action of these two
compartments is hence the prime candidate
for the realization of backpropagation in 
multi-layered networks
\citep{Bengio2014,Lee2015,Guerguiev2017}. 

\subsection{Learning Targets by Maximizing Correlation}

In the past, backpropagation algorithms 
for pyramidal neurons concentrated on 
learning rules that are explicitly dependent 
on an error term, typically the difference 
between top-down and bottom-up signals.
In this work, we considered an alternative 
approach. We postulate that the correlation 
between proximal and distal input constitutes
a viable objective function, which is to be
maximized in combination with homeostatic 
adaptation rules that keep proximal and 
distal inputs within desired working regimes.
Learning correlations between distinct
synaptic or compartmental inputs is
as a standard task for Hebbian-type learning,
which implies that the here proposed framework
is based not on supervised, but on biologically
viable unsupervised learning schemes.

The proximal input current $I_p$ is a linear 
projection of the proximal input space. 
Maximizing the correlation between $I_p$ and 
$I_d$ (the distal current), can therefore be regarded 
as a form of canonical correlation
analysis (CCA) \citep{Haerdle2007}. The idea 
of using CCA as a possible mode of 
synaptic learning has previously been investigated
by \citet{Haga2018}. Interestingly, according to 
the authors, a BCM-learning term in the plasticity 
dynamics accounts for a principal component analysis 
in the input space, while CCA requires an additional 
multiplicative term between local basal and apical 
activity. In contrast, our results indicate that 
such a multiplicative term is not required to drive 
basal synaptic plasticity towards a maximal alignment 
between basal and apical input, even in the presence 
of distracting principal components.
Apart from the advantage that this avoids the necessity 
of giving a biophysical interpretation of such cross-terms, 
it is also in line with the view that synaptic plasticity 
should be formulated in terms of local membrane voltage 
traces \citep{Clopath2010,Weissenberger2018}. 
According to this principle, distal compartments should 
therefore only implicitly affect plasticity in basal 
synapses, e.g.\ by facilitating spike initiation.

\subsection{Generalizability of the Model to Neuroanatomical
	Variabillity}

While some research on cortical circuits suggests the possibility of
generic and scalable principles that apply to different cortical
regions and their functionality \citep{Douglas2007,George2009,Larkum2013},
it is also well known that the anatomical properties of pyramidal neurons, in particular the dendritic structure, varies significantly across cortical regions 
\citep{Fuster1973,Funahashi1989}. More specifically, going from lower to higher areas of
the visual pathway, one can observe a significant increase of spines in the
basal dendritic tree \citep{Elston1997, Elston2000}, which can be associated with the fact that neurons in higher cortical areas generally encode more complex or even
multi-sensory information, requiring the integration of activity from a higher number
and potentially more distal neurons \citep{Elston2003,Luebke2017}.

With respect to a varying amount of basal synaptic inputs, it is interesting
to note that the dimensionality $N$ of the basal input patterns did not have
a large effect on the results of our model, see 
Fig.~\ref{fig_corr_dimension_scaling}--\ref{fig:classification_correlation},
as long as the homeostatic processes provided weight normalization.

Apart from variations in the number of spines, variability can also be observed within the dendritic structure itself 
\citep{Spruston2008,Ramaswamy2015}. Such differences obviously affect
the internal dynamics of the integration of synaptic inputs. 
Given the phenomenological nature of our neuron model, 
it is hard to predict how such differences would be reflected, given
the diverse dynamical properties that can arise from the dendritic structure
\citep{Hausser2000}.
The two models tested in our study can be regarded as two extreme cases,
where the point neuron represents a completely linear superposition of inputs 
and the compartment model being strongly nonlinear with respect to proximal
and distal inputs. In principle, pyramidal structures could also exhibit
properties in between, where the resulting plasticity processes would show
a mixture between the classical point neuron behavior (e.g.\ if a 
dimensionality reduction of the input via PCA is the main task) and a 
regime dominated by proximal-distal input correlations if top-down signals
should be predicted.

\subsection{Outlook}

Here we concentrated on one-dimensional distal inputs.
For the case of higher-dimensional distal input 
patterns, as for structured multi-layered networks,
it thus remains to be investigated how target signals 
are formed. However, as previous works have indicated, 
random top-down weights are generically sufficient 
for successful credit assignment and learning tasks
\citep{Lillicrap2016,Guerguiev2017}.
Therefore, we expect that our results can be
also transferred to deep network structures, 
for which plasticity is classically guided 
by local errors between top-down and bottom-up signals.

\section{Appendix}

\subsection{Alignment in the Classificaction Task}
\begin{figure}[t]
	\centering
	\includegraphics[width=1.0\columnwidth]{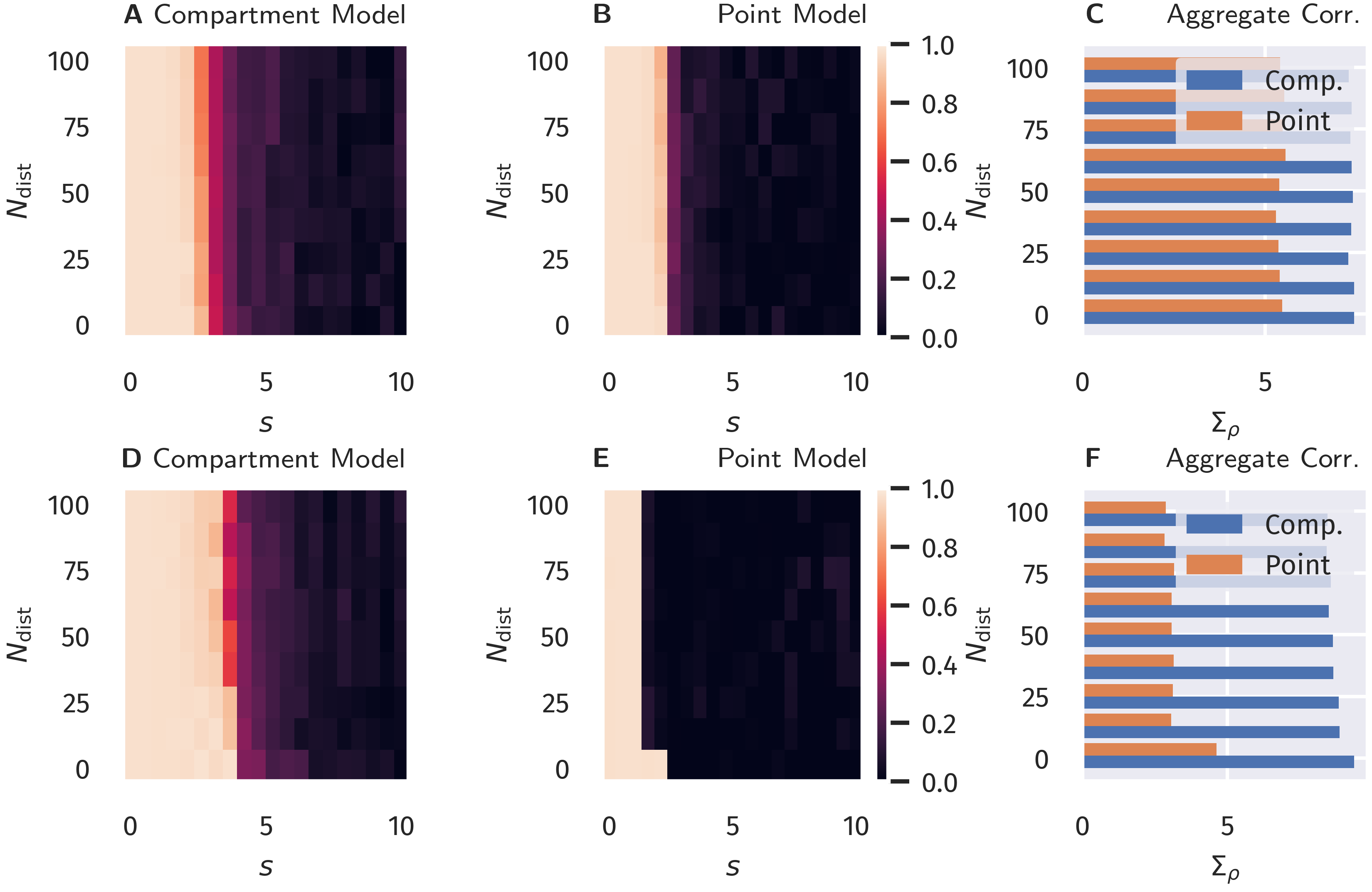}
	\caption{{\bf Alignment between Basal and Apical Input 
			after Binary Classification Learning.}
		Correlation between proximal and distal inputs
		after training, as described in Sect.~\ref{sect:classification}. 
		A--C: Classical Hebbian plasticity. 
		D--F: BCM rule.
		In the bar plot on the right the sum
		$\Sigma_{\rm acc}$ over $s=0,\,0.5,\,1.0\,..$ 
		of the results is shown as a function
		of $N_{\rm dist}$. Blue bars represents 
		the compartment model, orange the point model.}
	\label{fig:classification_correlation}
\end{figure}

Instead of measuring the model performance in the classification 
task presented in Sect.~\ref{sect:classification} by the fraction
of correctly classified patterns, 
as shown in Fig.~\ref{fig:classification_accuracy},
one can also use the correlation between $I_p$ and $I_d$, 
as done in Sect.~\ref{sect:alignment}. This is shown 
in Fig.~\ref{fig:classification_correlation}.
One observes a more pronounced difference between 
the point model and the compartment model, where 
the latter results in an overall better alignment 
for the tested parameter space.

\newpage
\subsection{Objective Function of BCM Learning in the Compartment Model}
\label{sect:Obj_Func}

\begin{figure}[t]
	\centering
	\includegraphics[width=0.9\columnwidth]{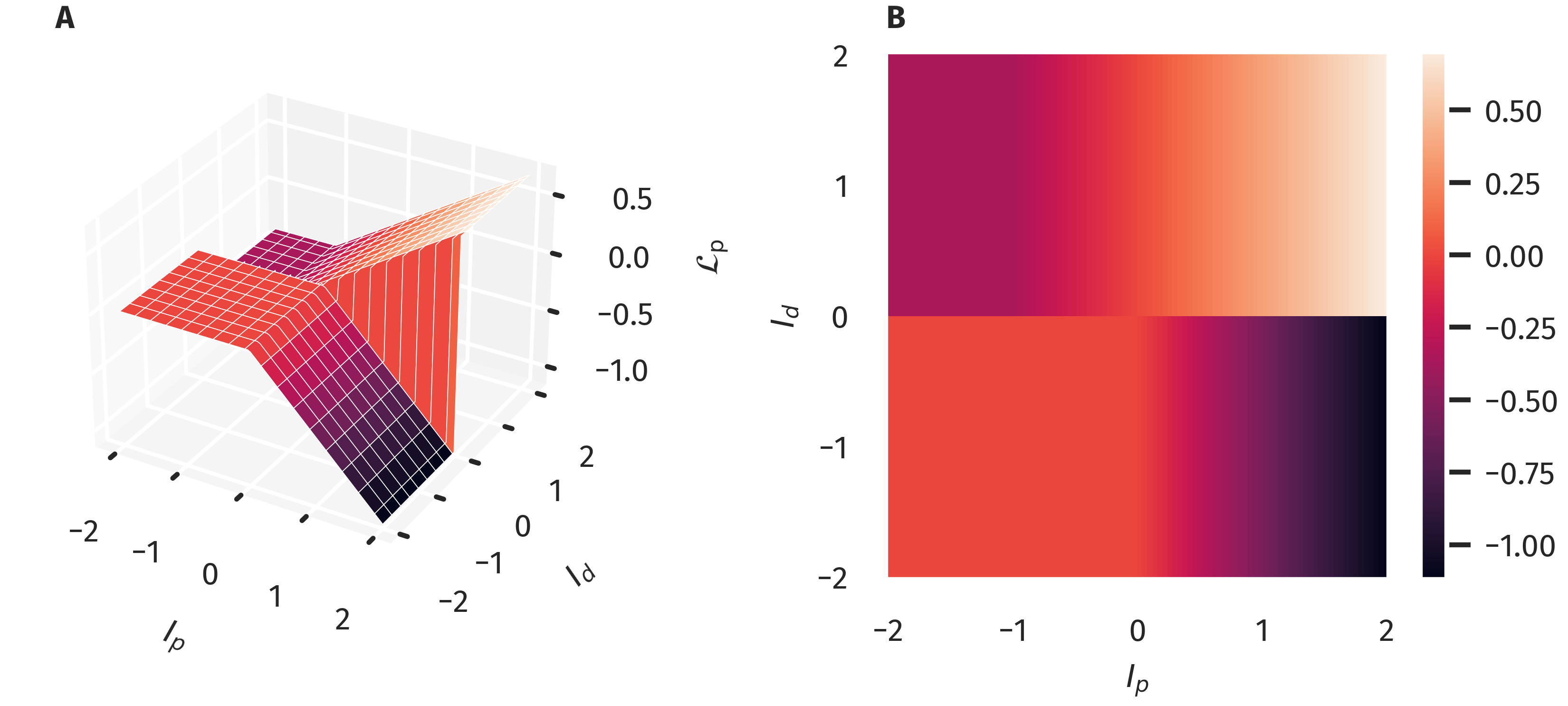}
	\caption{{\bf Objective Function for the Proximal Weight Update.} 
		The approximate objective function for the proximal 
		weights as given in \eqref{eq_obj_func} as a 3d-plot (A) and color-coded
		(B). This corresponds 
		to a combination of using \eqref{eq_comp_model} 
		together with \eqref{eq_bcm_rule}. Note the ridge-like
		structure along the $I_p$-$I_d$ diagonal, which 
		supports the alignment between proximal and distal input.}
	\label{fig:obj_func}
\end{figure}

To gain a better understanding of why the BCM-type 
learning rule in combination with the implemented 
compartment model drives the neuron towards 
the temporal alignment between $I_p$ and $I_d$, 
we can formalize the learning rule for the 
proximal weights in terms of an objective function.
For this purpose, we further simplify 
\eqref{eq_comp_model} by replacing the sigmoid 
functions $\sigma(x)$ by a simple step 
function $\Theta(x)$. This does not change the 
overall shape or topology of the activation
in the $(I_p,I_d)$ space but merely makes the 
smooth transitions sharp and instantaneous. 
Using $\Delta w_i \propto y\left(y - \theta_M \right) x_i$,
we find in this case
\begin{equation}
\Delta w_i \ \propto\  
\Big[ (1-\alpha) \Theta(I_d - \theta_{d})\Theta(p-\theta_{p1})
\\ + \alpha (\alpha - 1)\Theta(\theta_{d} - I_d)\Theta(p-\theta_{p0}) 
\Big]x_i \; .
\end{equation}
Noting that $\Theta(x)$ is the first derivative 
of the ReLu function $[x]^+ \equiv \max(0,x)$,
we find that this update rule can be written as
\begin{eqnarray}
\nonumber
\Delta w_i &\propto&
\frac{\partial \mathcal{L}_p}{\partial w_i}\\
\mathcal{L}_p &=& (1-\alpha) \Theta(I_d - 
\theta_{d})[p-\theta_{p1}]^+
+ \alpha (\alpha - 1)\Theta(\theta_{d} - I_d)[p-\theta_{p0}]^+ \; .
\label{eq_obj_func}
\end{eqnarray}
The objective function $\mathcal{L}_p$
is shown in Fig.~\ref{fig:obj_func}. 
One observes that states closer to 
the $I_p$-$I_d$ diagonal are preferred since
they tend to yield higher values of $\mathcal{L}_p$,
while the opposite is the case for off-diagonal 
states.

It should be noted, though, that the objective function 
is not scale-invariant (as 
would be e.g.\ if the squared error was used) in the 
sense that the prior distributions of both proximal 
and distal inputs need a certain mean and variance to 
cover a region of input states for which the described 
effects can take place. As a counterexample, one could 
imagine that the input samples only covered a flat area of 
$\mathcal{L}_p$, as for example in Fig.~\ref{fig:obj_func}B 
in the lower-left quadrant, leading to a zero average gradient. This 
is prevented, however, by the homeostatic processes 
acting simultaneously on the gains
and biases, making sure that the marginal distributions of $I_p$ and
$I_d$ are such that higher correlations are preferred. For example,
if we assume a Gaussian marginal distribution for both $I_p$ and
$I_d$ with zero means and a standard deviation of $0.5$ (which is
used as a homeostatic target in the simulations), the expected value
of $\mathcal{L}(I_p,I_d)$ is $-0.055$ if $I_p$ and
$I_d$ are completely uncorrelated, and $0.07$ in the perfectly correlated 
case.

\section*{Conflict of Interest Statement}

The authors declare that the research was conducted in the absence of any commercial or financial relationships that could be construed as a potential conflict of interest.

\section*{Author Contributions}

Both authors, F.S. and C.G., contributed equally to the
writing and review of the manuscript. F.S. provided the code,
ran the simulations, and prepared the figures.


\section*{Acknowledgments}

The authors acknowledge the financial support of
the German Research Foundation (DFG)

\section*{Data Availability Statement}
The simulation datasets for this study can be found under \url{https://cloud.itp.uni-frankfurt.de/s/mSRJ6BPXjwwHmfq}.
The simulation and plotting code for this project can be found under 
\url{https://github.com/FabianSchubert/frontiers_dendritic_coincidence_detection}.

\bibliographystyle{unsrtnat}
\bibliography{Manuscript.bib}  






\end{document}